\documentclass[twocolumn,aps,prd,superscriptaddress,floatfix,longbibliography,showkeys]{revtex4-1}  
\usepackage{xcolor}

\usepackage[draft]{changes}
\usepackage[normalem]{ulem}

\usepackage{graphicx}
\usepackage{amssymb} 
\usepackage{dcolumn}
\usepackage{bm}
\usepackage[mathlines]{lineno}
\usepackage[colorlinks,linkcolor=blue,anchorcolor=blue,citecolor=blue,urlcolor=blue]{hyperref}
\usepackage{multirow}
\usepackage{color}
\usepackage{amsmath}

\makeatletter
\renewcommand*{\@fnsymbol}[1]{\ensuremath{\ifcase#1\or \dagger\or *\or \ddagger\or
\mathsection\or \mathparagraph\or \|\or **\or \dagger\dagger \or
\ddagger\ddagger \else\@ctrerr\fi}} \makeatother

\usepackage{color}%

\usepackage[draft]{changes}
\usepackage[normalem]{ulem}
\definechangesauthor[name={gao}, color=blue]{gao}

\begin{document}
%
\title{Metavalent-Like Bonding Character Drives Giant Anharmonicity and Ultralow Thermal Conductivity in Rocksalt BeO}
%

%

%

\author{Xuejie Li}
\affiliation{State Key Laboratory for Mechanical Behavior of Materials, State Key Laboratory of Porous Metal Materials, School of Materials Science and Engineering, Xi'an Jiaotong University, Xi'an 710049, China}

\author{Shengying Yue}
\affiliation{Laboratory for Multiscale Mechanics and Medical Science, State Key Laboratory for Strength and Vibration of Mechanical Structures, School of Aerospace, Xi’an Jiaotong University, Xi'an 710049, China}


\author{Bolin Wang}
\affiliation{China Mobile Group Shaanxi Co., Ltd. Xi’an 710077, China}

\author{Xiaolong Yang (corresponding author)}
\email[E-mail: ]{yangxl@cqu.edu.cn}
\affiliation{College of Physics, Chongqing Key Laboratory for Strongly Coupled Physics, Chongqing University, Chongqing 401331, China}
\affiliation{Center of Quantum Materials and Devices, Chongqing University, Chongqing 401331, China}

\author{Turab Lookman}
\affiliation{State Key Laboratory for Mechanical Behavior of Materials, State Key Laboratory of Porous Metal Materials, School of Materials Science and Engineering, Xi'an Jiaotong University, Xi'an 710049, China}

\author{Jun Sun}
\affiliation{State Key Laboratory for Mechanical Behavior of Materials, State Key Laboratory of Porous Metal Materials, School of Materials Science and Engineering, Xi'an Jiaotong University, Xi'an 710049, China}

\author{Xiangdong Ding (corresponding author)}
\email[E-mail: ]{dingxd@mail.xjtu.edu.cn}
\affiliation{State Key Laboratory for Mechanical Behavior of Materials, State Key Laboratory of Porous Metal Materials, School of Materials Science and Engineering, Xi'an Jiaotong University, Xi'an 710049, China}

\author{Zhibin Gao (corresponding author)}
\email[E-mail: ]{zhibin.gao@xjtu.edu.cn}
\affiliation{State Key Laboratory for Mechanical Behavior of Materials, State Key Laboratory of Porous Metal Materials, School of Materials Science and Engineering, Xi'an Jiaotong University, Xi'an 710049, China}

\date{\today}

\begin{abstract}
\textbf{ABSTRACT}\par
\vspace{0.5em}
Identifying materials with intrinsically low lattice thermal conductivity ($\kappa_L$) underpins thermoelectric and phase-change technologies, yet simple structural descriptors often fail to capture how chemical bonding controls lattice dynamics. Here we compare thermal transport in two BeO polymorphs of identical chemistry and atomic mass: covalent zincblende (zb) and metavalent-like rocksalt (rs). Despite being denser, rs-BeO exhibits a $\kappa_L$ more than an order of magnitude below zb-BeO (24 vs. 357 W m$^{-1}$ K$^{-1}$ at 300 K), breaking the conventional density-$\kappa_L$ correlation. $\emph{Ab initio}$ bonding analysis traces this anomaly to metavalent-like bonding in rs-BeO, which produces electron delocalization, long-range interatomic force constants, and giant quartic anharmonicity that drives strong four-phonon scattering, whereas stiff covalent bonding sustains efficient heat conduction in zb-BeO. Quantitative modeling of the rocksalt phase requires explicit phonon renormalization and four-phonon scattering, effects negligible in zb-BeO. Bonding character, rather than density, governs thermal transport in this ultra-wide-bandgap oxide.
\end{abstract}

\begingroup
\normalsize
\endgroup



\maketitle

Beryllium oxide (BeO) is the only alkaline-earth oxide that crystallizes in the wurtzite rather than the rocksalt structure. Its low atomic mass, high Debye temperature, exceptional hardness, high melting point, and outstanding thermal conductivity combined with high electrical resistivity~\cite{10.1063/1.3075814,Ivanovskii2009,Sorokin2006,PhysRevB.76.085407,10.1063/1.336756,10.1063/1.1659844,Vidal-Valat:a26524,10.1063/1.1659844,Qi-Li_2008,ROESSLER1969157,Wdowik_2010} make it attractive for optics, power electronics, reactor moderators, and chip-carrier substrates~\cite{10.1063/1.3359706,Wdowik_2010}. Most studies have addressed wurtzite BeO (w-BeO)~\cite{10.1063/1.1659844,https://doi.org/10.1111/j.1151-2916.1966.tb15389.x,10.1063/1.1709787,doi:10.1021/acsomega.9b00174,PhysRevB.77.224303,MALAKKAL201779,Vidal-Valat:a26524,Milman_2001}, whereas the zincblende (zb-BeO) and rocksalt (rs-BeO) polymorphs have received far less attention.



Recent study reported that rs-BeO possesses an extremely high static dielectric constant (approximately 271$\epsilon_0$, where $\epsilon_0$ is the vacuum permittivity), arising from strong Coulomb repulsion of the electron cloud that elongates the Be-O bond and softens the transverse optical (TO) phonon~\cite{Yim2015,Cao2024}. This soft mode signals structural instability and suggests a route toward new ferroelectric devices, yet the thermal transport of rs-BeO and its connection to this unusual bonding character remain unexplored.

Notably, the properties of rs-BeO such as a soft TO mode, enhanced Born effective charges, a very large dielectric constant, and a large Gr\"uneisen parameter, closely resemble those of metavalent solids, a class extensively studied by Wuttig et al. for their phase-change and thermoelectric performance~\cite{https://doi.org/10.1002/adma.201803777,Shportko2008,https://doi.org/10.1002/pssb.200982010,doi:10.1021/acs.nanolett.9b03435,https://doi.org/10.1002/adma.201806280,https://doi.org/10.1002/adma.202208485,https://doi.org/10.1002/advs.202308578,https://doi.org/10.1002/adma.201801787,https://doi.org/10.1002/adfm.201904862}. Such anomalies resist a conventional ionic or covalent description and instead point to electronic states that are neither strongly localized nor fully delocalized, a situation known to produce pronounced anharmonicity and atypical transport properties.



These parallels raise a basic question: can metavalent-like bonding, so far confined to narrow-gap chalcogenides, arise in an ultra-wide-bandgap insulator (E$_g>$10.0 eV) with negligible electrical conductivity, and if so, how does it govern thermal transport? We address this through a controlled first-principles comparison of zb-BeO and rs-BeO, two phases of identical chemistry and atomic mass but contrasting bonding character. Using pCOHP, ICOBI, ELF, charge-density response, and distance-resolved interatomic force constants, we show that rs-BeO is a $p$-orbital-dominated system exhibiting metavalent-like bonding characteristics, marked by weakened two-center Be-O bonding and enhanced bond delocalization, whereas zb-BeO displays conventional $p$-bonded covalent behavior.

To evaluate $\kappa_L$, we employ anharmonic phonon renormalization within the self-consistent phonon (SCPH) framework, explicit three- and four-phonon scattering, and off-diagonal coherent transport contributions~\cite{li2026four,wang2023role,hao2025copper}. The long-range interactions associated with metavalent-like bonding in rs-BeO generate strong lattice anharmonicity and suppress $\kappa_L$ by more than an order of magnitude relative to zb-BeO, demonstrating that bonding character, rather than density, governs thermal transport. Here, zb-BeO and rs-BeO are selected because they have identical chemical composition and atomic masses while both possessing cubic symmetry, allowing the effect of bonding topology to be isolated without additional crystallographic anisotropy. In contrast, w-BeO has hexagonal symmetry and would introduce anisotropy as an additional variable~\cite{10.1063/1.5040887,10.1063/5.0201393}.

\begin{figure*}
\includegraphics[width=1.55\columnwidth]{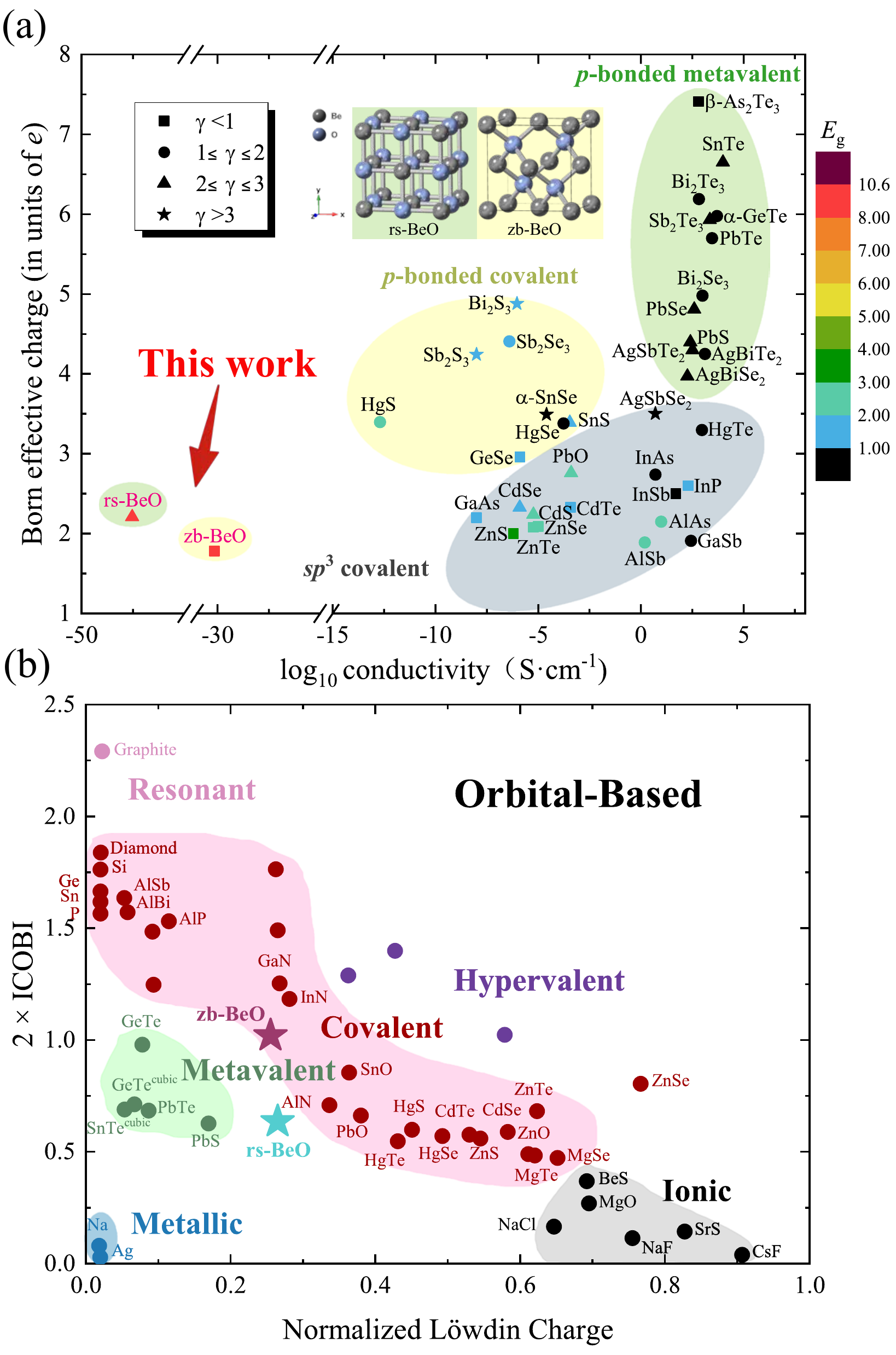}
\caption{%
Classification of semiconductors by different bonding descriptors. (a) Electrical conductivity and Born effective charge, with band gap ($E_g$) indicated by color (black:$\textless$1 eV; blue:1-2 eV; light green:2-3 eV; dark green:3-8 eV; red:$\textgreater$8 eV) and Gr\"{u}neisen parameter ($\gamma$) indicated by symbol shape (square:$\gamma$$\textless$1; circle:1$\leq$$\gamma$$\leq$2; triangle:2$\leq$$\gamma$$\leq$3; star:$\gamma$$\textgreater$3). The green, yellow, and grey ellipses denote the regions of $p$-bonded metavalent materials, $p$-bonded covalent systems, and \emph{sp$^3$} covalent crystals, respectively. Insets show the crystal structures of zb-BeO and rs-BeO, where grey and blue spheres represent Be and O atoms. All data are provided in Table S1 in the Supplementary Material. (b) Orbital-based bonding map constructed from the normalized L\"{o}wdin charge and the electron-sharing descriptor $2\times$ICOBI. The horizontal axis represents electron transfer, defined by the normalized L\"{o}wdin charge, whereas the vertical axis represents electron sharing, evaluated as twice the integrated crystal orbital bond index. Conventional covalent, metavalent, resonant, metallic, hypervalent, and ionic bonding regions are shown as background guides. The zb-BeO and rs-BeO points are calculated in this work and are highlighted by star symbols. Except for the BeO data, the reference data in panel (b) are taken from Ref.~\cite{https://doi.org/10.1002/advs.202308578}. Adapted from ref.~\cite{https://doi.org/10.1002/advs.202308578}. Copyright 2023, The Authors, Advanced Science published by Wiley-VCH GmbH under a Creative Commons Attribution 4.0 International License (http://creativecommons.org/licenses/by/4.0/).
\label{fig1}}
\end{figure*}

\begin{figure*}
\includegraphics[width=2.0\columnwidth]{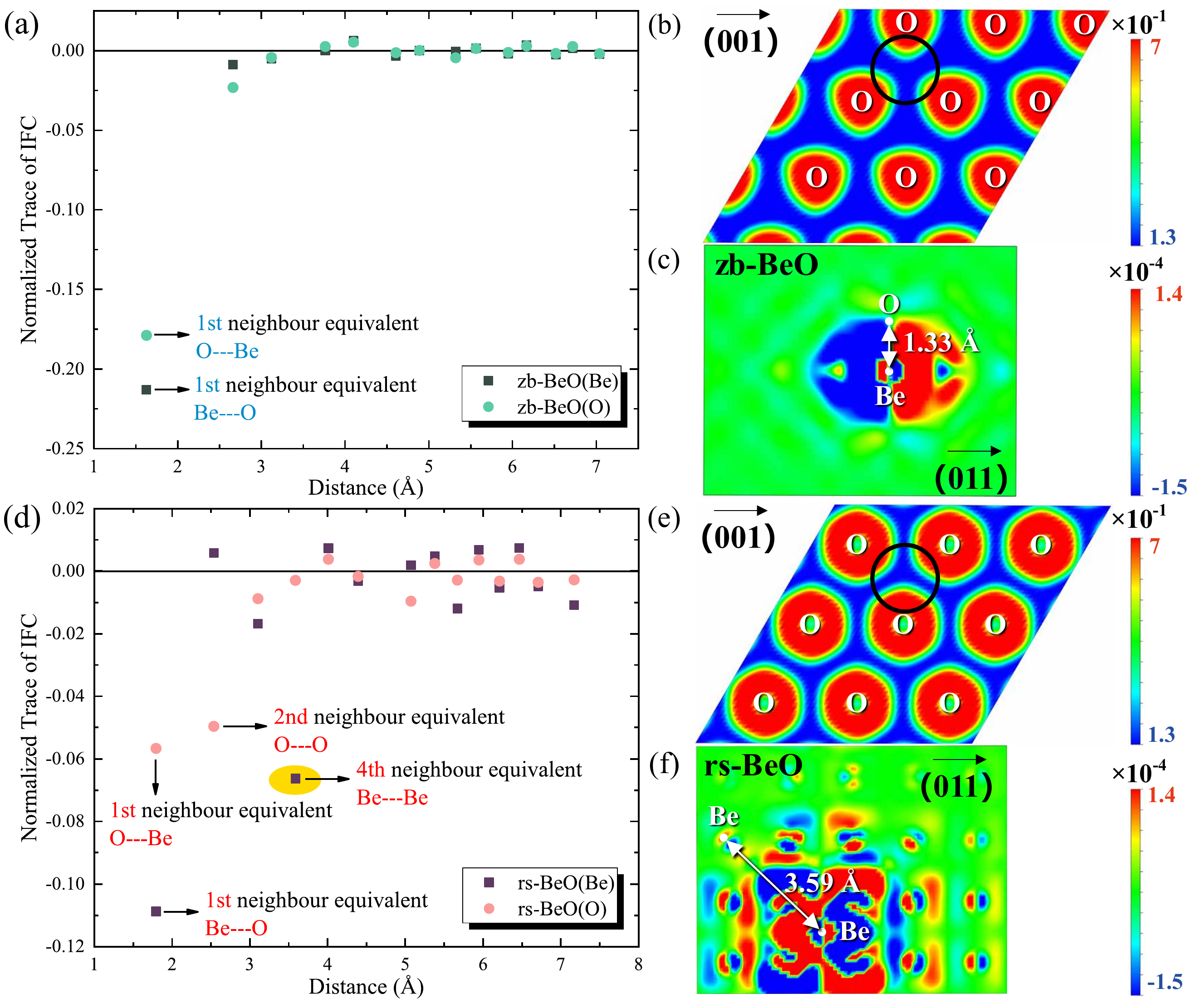}
\caption{
Bonding characteristics of zb-BeO and rs-BeO. (a,d) Normalized traces of interatomic force-constant (IFC) tensors as a function of interatomic distance for zb-BeO and rs-BeO, respectively. Different colors denote IFCs with a specific element chosen as the origin atom. (b,e) Equilibrium electron-density distributions in the Be atomic layer for zb-BeO and rs-BeO. Panels (b) and (e) are plotted using the same color scale. Black circles highlight the interatomic regions where notable differences in electron delocalization are observed between the two phases. (c,f) Charge-density distortions induced by a small displacement of the central Be atom in zb-BeO and rs-BeO. Panels (c) and (f) are plotted using the same color scale. Color bars in (b,e) and (c,f) indicate the electron density and charge-density variation, respectively, in units of e/$\text{\AA}^3$.
\label{fig2}} 
\end{figure*}

\begin{figure*}
\includegraphics[width=2.0\columnwidth]{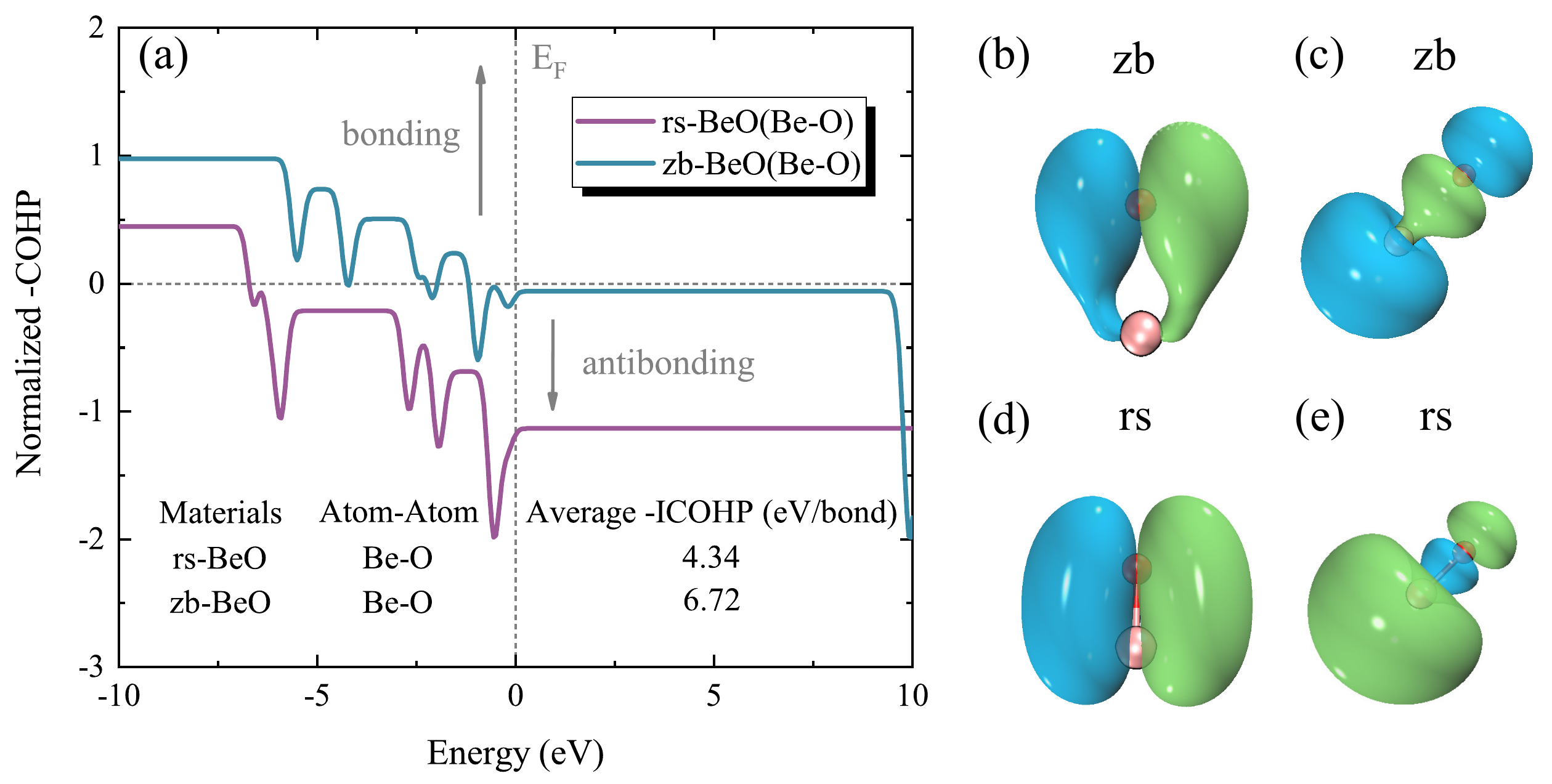}
\caption{
Projected crystal orbital Hamilton population (COHP) analysis for the Be-O interactions in zb-BeO and rs-BeO, calculated using the LOBSTER~\cite{https://doi.org/10.1002/jcc.26353}. (a) Energy-resolved -COHP curves for Be-O bonds in both phases. Positive -COHP values indicate bonding states, while negative values correspond to antibonding states. All energies are referenced to the Fermi level (E$_F$=0 eV). The inset shows the enlarged near-Fermi-energy region. The reported -ICOHP values are obtained by integrating the -ICOHP up to E $_F$  and averaging over nearest-neighbor Be-O bonds, and are given in units of eV per bond. (b,c) Valence-band-edge and conduction-band-edge crystal orbitals associated with Be-O bonding in zb-BeO. (d,e) Corresponding crystal orbitals in rs-BeO.
\label{fig3}}
\end{figure*}


Wuttig et al.~\cite{https://doi.org/10.1002/adma.201803777} have shown that phase-change materials (PCMs) and related compounds exhibit a distinctive bonding mechanism intermediate between covalent and metallic states, termed ``metavalent bonding''. From a quantum-mechanical perspective, the ground state of such systems is a superposition of symmetry-equivalent saturated bond configurations~\cite{PhysRevB.8.660}, with electrons delocalized over all possible bonding arrangements. This electronic structure, observed in group IV–VI, V$_2$-VI$_3$, and group V compounds, gives rise to very low $\kappa_L$. The long-range nature of these interactions produces a flat anharmonic potential energy surface, enabling large atomic fluctuations and enhanced structural entropy that facilitate rapid phase transitions~\cite{hao2026llm}. These features underpin the exceptional performance of metavalently bonded materials in both thermoelectric and phase-change applications~\cite{https://doi.org/10.1002/adma.201803777,Shportko2008,https://doi.org/10.1002/pssb.200982010,doi:10.1021/acs.nanolett.9b03435,https://doi.org/10.1002/adma.201806280,https://doi.org/10.1002/adma.202208485,https://doi.org/10.1002/advs.202308578,https://doi.org/10.1002/adma.201801787,https://doi.org/10.1002/adfm.201904862}.


Materials with metavalent bonding typically display several characteristic features: (i) bonding dominated by the $\emph{p}$-electrons, (ii) octahedral coordination, and (iii) pronounced structural tunability~\cite{PBLittlewood_1980,doi:10.1126/science.aad8688,10.1147/rd.83.0215}, arising from high sensitivity to temperature, pressure~\cite{CHATTOPADHYAY1986879}, and stoichiometry~\cite{Lencer2008}. These systems also exhibit distinctive dynamical properties, including soft transverse optical (TO) phonon modes, anomalously large Born effective charges, and high dielectric constants—all rooted in their unique electronic structure and bonding character. Recent work on rs-BeO~\cite{Cao2024} has shown that strong electron cloud overlap and Coulomb repulsion elongate the Be-O bond and soften the TO phonons. Given its flexible rocksalt structure, the group-VI oxygen character, and the unexpectedly large static dielectric constant, we hypothesize that rs-BeO may exhibit metavalent-like bonding features. To test this hypothesis, we carried out systematic $\emph{Ab initio}$ calculations.


Following the classification criteria proposed by Wuttig et al.~\cite{https://doi.org/10.1002/adma.201803777}, we analyze the bonding character of both BeO polymorphs. Fig.~\ref{fig1}(a) maps materials by plotting electrical conductivity versus Born effective charge. The color coding distinguishes $\emph{p}$-bonded metavalent systems (green), $\emph{p}$-bonded covalent compounds (yellow), and \emph{sp$^3$} covalent semiconductors (gray). Our results place rs-BeO near the metavalent-like region, while zb-BeO falls into the conventional covalent category. Comparing the Gr\"{u}neisen parameter ($\gamma$) further differentiates the two phases: most metavalent materials, including rs-BeO ($\gamma$=2-3), exhibit strong anharmonicity typically associated with low $\kappa_L$, whereas zb-BeO shows a much smaller value ($\gamma$=0.94). The band gap analysis reveals that most known metavalent systems have gaps below 3 eV (ZnS at 3.76 eV represents the upper limit in prior studies). However, both rs-BeO (E$_g$=10.6 eV) and zb-BeO (E$_g$=9.3 eV) far exceed this range, demonstrating that metavalent-like bonding features can appear in wide-gap systems. This finding significantly extends the known domain of this metavalent-like bonding mechanism.

To further quantify the bonding character, Fig.~\ref{fig1}(b) presents an orbital-based bonding map using the normalized L\"{o}wdin charge as a measure of electron transfer (ET) and $2\times$ICOBI as a measure of electron sharing (ES)~\cite{https://doi.org/10.1002/advs.202308578}. In this representation, conventional covalent materials are characterized by relatively large ES and small-to-moderate ET, whereas ionic compounds exhibit large ET but weak ES. Metavalent systems occupy an intermediate regime with reduced two-center electron sharing compared with normal covalent solids, but without reaching the strongly ionic limit. The calculated zb-BeO point lies close to the covalent region, consistent with its localized Be--O bonding and large ICOBI value. In contrast, rs-BeO shows a clear reduction in ES while maintaining a comparable normalized L\"{o}wdin charge, shifting it toward the metavalent region. This contrast indicates that the rocksalt phase does not simply represent a conventional covalent or ionic bonding limit, but instead exhibits metavalent-like bonding characteristics with weakened two-center electron sharing and enhanced bonding delocalization. This orbital-based descriptor therefore provides an independent quantitative support for the assignment of metavalent-like bonding in rs-BeO.


To elucidate the bonding differences between the two phases, we examine the distance-dependent second-order interatomic force constants (IFCs) shown in Fig.~\ref{fig2}. Following the method of Lee et al.~\cite{Lee2014}, we normalize the IFCs tensor traces to the self-interaction terms and compute the corresponding 2-norms; both approaches yield consistent trends. Both structures exhibit non-monotonic IFC behavior, in contrast to conventional covalent or ionic systems where interaction strength decays smoothly with distance.


In rs-BeO, strong long-range interactions are evident at specific coordination shells, most notably at the fourth nearest neighbors. The fourth-neighbor coupling is unexpectedly strong despite the greater interatomic distance compared to the second and third neighbors, which lie outside the metavalent network. The metavalent-like bonding also produces anomalous positive IFC values (corresponding to negative spring constants) at the seventh, ninth, and fourteenth neighbor shells that is a characteristic signature of long-range electronic coupling in metavalent systems~\cite{Lee2014}. These distinctive IFC features directly underpin the contrasting thermal transport behaviors of the two phases, as discussed in Section II C.


To assess the electron delocalization associated with metavalent-like bonding, Figs.~\ref{fig2}(b) and (e) show the equilibrium charge density distributions. Metavalent bonds typically produce high charge density between atoms, reflecting strong electron sharing~\cite{Lee2014,PhysRevLett.107.175503}. As shown, rs-BeO exhibits more delocalized charge density between Be atoms than zb-BeO, consistent with enhanced bonding interactions and the known behavior of metavalent solids~\cite{YUE201889}. Figs.~\ref{fig2}(c) and (f) display the charge density response to a small displacement of the central Be atom, capturing fourth-nearest-neighbor interactions in the (011) plane. Under identical displacement conditions, rs-BeO shows stronger and more spatially extended charge redistribution than zb-BeO, directly indicating enhanced metavalent bonding with strong interatomic electronic coupling. We note that these results are robust with respect to the choice of exchange-correlation functionals, as consistent trends are obtained from additional calculations using both PBE and PBEsol, as shown in Figs. S2 and S3 in the Supplemental Material.

To further characterize the bonding nature, we analyze the projected crystal orbital Hamilton population (-COHP), as shown in Fig.~\ref{fig3}(a). In this framework, positive values of -COHP correspond to bonding states, while negative values indicate antibonding states, with all energies referenced to the Fermi level (E$_F$=0 eV)~\cite{doi:10.1126/sciadv.abg1449,10.1063/5.0271507}.

In zb-BeO, pronounced bonding contributions are observed deep in the valence band, with a clear energetic separation between bonding and antibonding states. The average -ICOHP over all nearest-neighbor Be-O bonds is 6.72 eV per bond, indicating strong directional covalent interactions. The occupied states are dominated by well-localized bonding character, while antibonding states remain largely unoccupied above the Fermi level, consistent with a conventional localized bonding picture.

In contrast, rs-BeO exhibits significantly reduced -COHP magnitudes and a weaker energy separation between bonding and antibonding states. The average -ICOHP decreases to 4.34 eV per bond, reflecting a substantial weakening of conventional two-center covalent bonding. Notably, bonding and antibonding contributions become partially mixed near the Fermi level, indicating incomplete orbital separation and enhanced electronic delocalization. This behavior suggests a deviation from an ideal localized bonding picture toward a more delocalized bonding regime.

This trend is consistent with complementary electronic structure analyses. The band structure and density of states (Figs.~S7 and S8 in the Supplemental Material) show strong O-$p$–Be-$s$ hybridization in zb-BeO, characteristic of localized covalent bonding. In rs-BeO, the hybridization is weaker and more distributed in energy, accompanied by more extended conduction states, consistent with enhanced electronic delocalization.

The real-space orbital analysis further reinforces this distinction. The highest occupied crystal orbitals (HOMOs) in zb-BeO [Figs.~\ref{fig3}(b) and (d)] exhibit strong localization along Be-O bonds, with charge density concentrated between nearest-neighbor atoms, indicative of directional covalent bonding. In contrast, the lowest unoccupied crystal orbitals (LUMOs) in rs-BeO [Figs.~\ref{fig3}(c) and (e)] display spatially extended distributions spanning multiple atomic sites, reflecting enhanced long-range electronic coupling.

Importantly, this electronic delocalization is consistent with the presence of long-range interatomic force constants and enhanced anharmonic interactions discussed in Fig.~\ref{fig2}. Taken together, these results indicate that the bonding in rs-BeO cannot be fully described within a purely localized two-center covalent framework, but instead exhibits characteristics of a more delocalized bonding regime with mixed ionic–covalent features, which is consistent with the observed suppression of lattice thermal transport.

\begin{figure*}
\includegraphics[width=2.0\columnwidth]{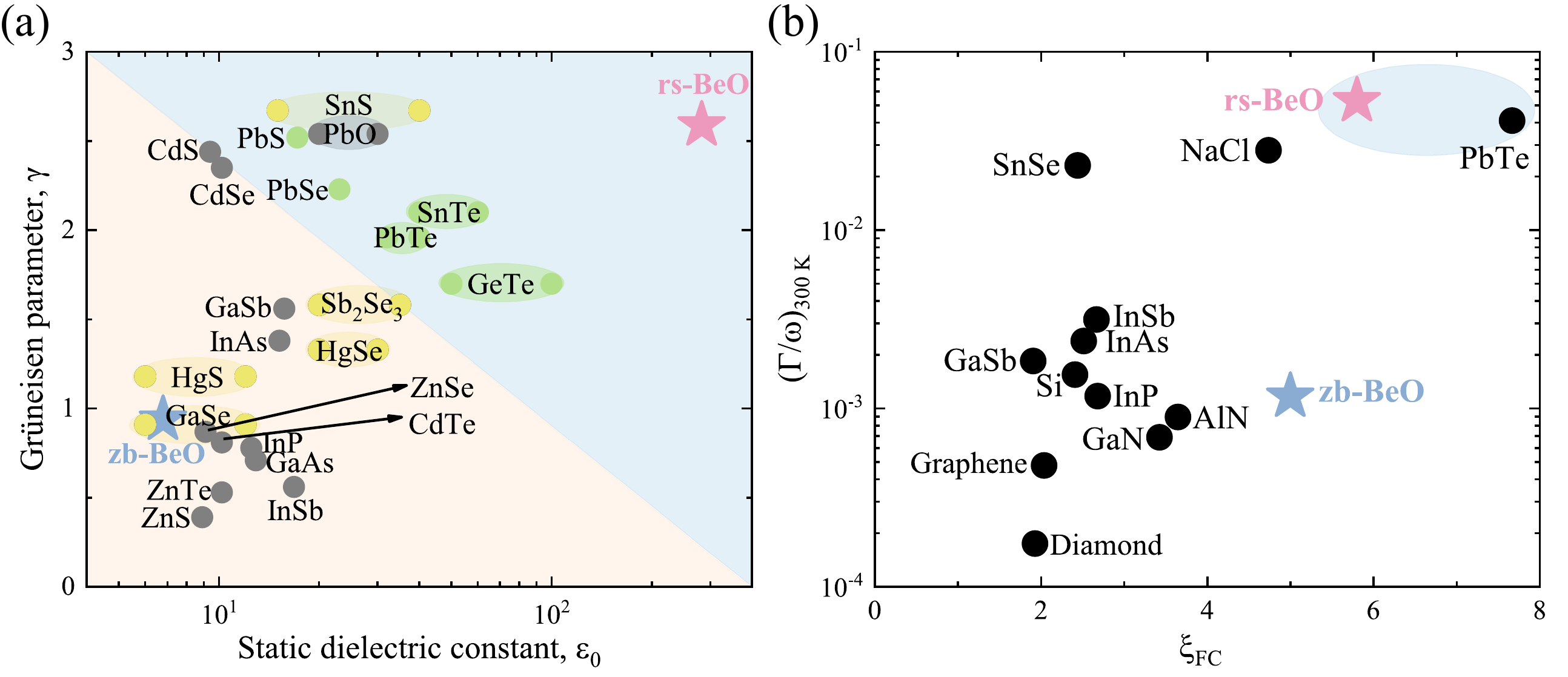}
\caption{
Bonding-related response descriptors for representative materials and BeO polymorphs. (a) Relationship between the static dielectric constant ($\varepsilon_0$) and the Gr\"{u}neisen parameter ($\gamma$). The background shading serves as a guide to the eye, distinguishing materials with relatively localized bonding and weak dielectric response from those with enhanced dielectric response and stronger anharmonicity. Representative covalent, polar covalent, ionic, and metavalent-like compounds are included to provide a comparative response-property landscape. The numerical data used in panel (a), including literature values of ($\varepsilon_0$) and ($\gamma$), are provided in the Supplementary Material. (b) Comparison between the normalized IFC decay length $\xi_{\mathrm{FC}}$ and the low-frequency phonon linewidth-to-frequency ratio at 300 K. Here, $\xi_{\mathrm{FC}}$ = $\xi_{\mathrm{fit}}$/$d_{NN}$ , where $\xi_{\mathrm{fit}}$ is extracted from an exponential fit to the shell-averaged real-space IFC envelope and $d_{NN}$ is the nearest-neighbor distance. The damping descriptor $(\Gamma/\omega)_{300\,\mathrm{K}}$, is evaluated for low-frequency heat-carrying modes with ($\omega_{\lambda} < 0.3\,\omega_{\max}$). The BeO polymorphs are highlighted by star symbols. Compared with zb-BeO, rs-BeO exhibits a longer normalized IFC decay length and a much larger phonon damping ratio, indicating that the metavalent-like bonding response in rs-BeO is accompanied by more extended force interactions and substantially enhanced anharmonic phonon scattering.
\label{fig4}}
\end{figure*}


Building on the bonding analysis, we examine bonding-related descriptors that connect dielectric response, anharmonicity, force-constant range, and phonon damping, as shown in Fig.~\ref{fig4}. To further substantiate the metavalent-like bonding assignment, we compare rs-BeO and zb-BeO with representative materials using established bonding-related descriptors, including the static dielectric constant $\varepsilon_0$ and the Gr\"{u}neisen parameter $\gamma$. This map is useful because a large dielectric response reflects strong polarizability and enhanced electronic screening, whereas a large $\gamma$ indicates strong sensitivity of phonon frequencies to volume or strain. Conventional covalent materials are mainly located in the low-$\varepsilon_0$/low-$\gamma$ region, while typical metavalent and highly polarizable compounds move toward larger $\varepsilon_0$ and enhanced anharmonicity. The rs-BeO point lies far from zb-BeO, exhibiting both an exceptionally large static dielectric constant and a large Gr\"{u}neisen parameter. This behavior indicates that the rocksalt phase combines strong dielectric response with pronounced lattice anharmonicity, consistent with metavalent-like bonding. 

Fig.~\ref{fig4}(b) further correlates the normalized interatomic force constants (IFCs) decay length
$\xi_{\mathrm{FC}}$ with the low-frequency phonon linewidth-to-frequency
ratio at 300~K. The normalized IFC decay length is defined as
\begin{equation}
    \xi_{\mathrm{FC}}=\frac{\xi_{\mathrm{fit}}}{d_{\mathrm{NN}}},
\end{equation}
where $d_{\mathrm{NN}}$ is the nearest-neighbor distance and
$\xi_{\mathrm{fit}}$ characterizes the spatial decay of the harmonic IFCs. 
Specifically, $\xi_{\mathrm{fit}}$
is obtained by fitting the shell-averaged Frobenius norm of the
off-diagonal harmonic IFC blocks to an exponential envelope,
\begin{equation}
    \bar{A}(r)=A_{0}\exp\left(-\frac{r}{\xi_{\mathrm{fit}}}\right),
\end{equation}
with
\begin{equation}
    \bar{A}(r)
    =
    \left\langle
    \left\|\boldsymbol{\Phi}^{(2)}_{ij}\right\|_{\mathrm{F}}
    \right\rangle_{\left|\mathbf{r}_{i}-\mathbf{r}_{j}\right|=r},
    \qquad i\neq j,
\end{equation}
where $\boldsymbol{\Phi}^{(2)}_{ij}$ denotes the harmonic IFC tensor
block between atoms $i$ and $j$, and
$\left\|\boldsymbol{\Phi}^{(2)}_{ij}\right\|_{\mathrm{F}}$ is its
Frobenius norm. The phonon-damping descriptor
$(\Gamma/\omega)_{300\,\mathrm{K}}$ is evaluated at 300~K for
low-frequency heat-carrying modes satisfying
$\omega_{\lambda}<0.3\,\omega_{\max}$. This comparison therefore links
the spatial persistence of harmonic force interactions to phonon
damping. Compared with zb-BeO, rs-BeO exhibits both a longer normalized
IFC decay length and a much larger low-frequency linewidth-to-frequency
ratio, indicating that the more extended force-constant network in the
rocksalt phase is accompanied by substantially stronger anharmonic
phonon scattering.

\begin{figure*}
\includegraphics[width=2.0\columnwidth]{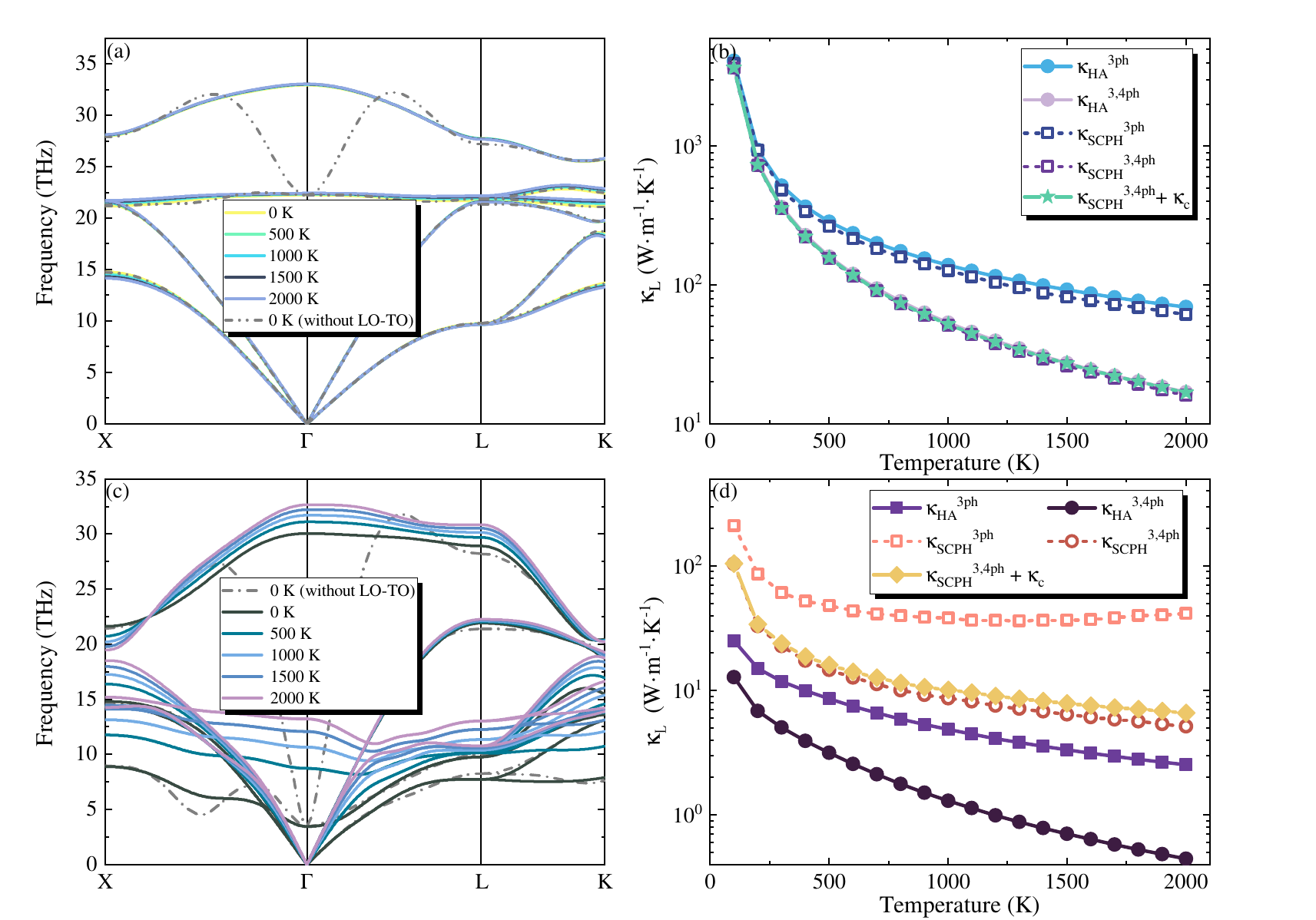}
\caption{
Temperature-dependent phonon dispersions and lattice thermal conductivity of zb-BeO and rs-BeO. (a, c) Temperature-dependent phonon dispersions of zb-BeO and rs-BeO calculated within the SCPH framework. The grey dash-dotted lines show the 0 K dispersions without LO-TO splitting. (b, d) Temperature-dependent lattice thermal conductivity $\kappa_L$ of zb-BeO and rs-BeO calculated at different levels of theory. $\kappa_{\mathrm{HA}}^{\mathrm{3ph}}$ and $\kappa_{\mathrm{HA}}^{\mathrm{3,4ph}}$ denote harmonic-phonon results with three-phonon scattering only and with both three- and four-phonon scattering, respectively. $\kappa_{\mathrm{SCPH}}^{\mathrm{3ph}}$ and $\kappa_{\mathrm{SCPH}}^{\mathrm{3,4ph}}$ denote the corresponding results obtained using SCPH-renormalized phonons. $\kappa_{\mathrm{c}}$ represents the coherent phonon contribution, and the total $\kappa_L$ is given by $\kappa_{\mathrm{SCPH}}^{\mathrm{3,4ph}}+\kappa_{\mathrm{c}}$.
\label{fig5}}
\end{figure*}

We then examine the anharmonic lattice dynamics of zb-BeO and rs-BeO from 0 to 2000 K within the self-consistent phonon (SCPH) framework. The calculated phonon dispersions are shown in Figs.~\ref{fig5}(a) and (c). The absence of imaginary frequencies confirms the dynamical stability of both phases, and this conclusion is robust with respect to the choice of exchange-correlation functionals, as additional calculations using both PBE and PBEsol also show no imaginary modes, particularly for rs-BeO (see Fig.~S1 in the Supplemental Material), and their thermodynamic stability is further supported by negative formation energies of -7.02~eV/f.u. for zb-BeO and -6.15~eV/f.u. for rs-BeO.

Although rs-BeO is not the equilibrium structure under ambient conditions, it is an experimentally established high-pressure phase rather than a hypothetical polymorph: X-ray diffraction has demonstrated the wurtzite-to-rocksalt (B1) transition at $137\pm5$~GPa, with the pure B1 phase observed up to 175~GPa~\cite{https://doi.org/10.1002/pssb.200405247}, consistent with earlier Raman work indicating a kinetically sluggish transition requiring substantial overpressure~\cite{PhysRevB.31.8332}. The open question is therefore not whether rs-BeO can be synthesized, but whether it can be retained upon decompression. Since the rocksalt phase lies approximately 0.483 eV per atom above the ambient wurtzite phase~\cite{Yim2015}, we do not claim that it can be readily recovered or synthesized at ambient pressure. Its ambient-pressure recoverability remains experimentally unestablished. We therefore treat rs-BeO as an experimentally realized high-pressure polymorph and a high-energy reference structure for isolating how rocksalt bonding topology affects anharmonicity and thermal transport.

Its finite-temperature dynamical stability is supported by phonon dispersions over a broad pressure range, temperature-dependent thermodynamic properties (Figs. S4-S6), and AIMD simulations at 0 and 200 GPa from 300 to 2000 K, which show bounded energy fluctuations and well-defined Be-O, Be-Be, and O-O correlation peaks (Figs. S15-S18). These results establish harmonic phonon stability and finite-temperature dynamical stability, which should be distinguished from thermodynamic stability or experimental recoverability at ambient pressure. Examples of other high-pressure phases recovered metastably at ambient conditions are given in the Supplementary Material.


For zb-BeO, the phonon dispersions change very little with temperature. The acoustic branches remain nearly unchanged, and the optical branches shift only slightly. A clear LO-TO splitting appears near the $\Gamma$ point, arising from the partially ionic Be-O bonds and the associated long-range Coulomb interaction. This splitting remains well defined and essentially temperature-independent, indicating that the polar bonding character of zb-BeO is robust.


In contrast, rs-BeO exhibits pronounced phonon renormalization. Several optical branches show clear frequency hardening with increasing temperature, and the intermediate-frequency modes also shift substantially. These features indicate much stronger anharmonic effects, particularly higher-order (quartic) contributions. Compared with zb-BeO, the temperature dependence of both the optical branches and the LO-TO splitting in rs-BeO is more strongly influenced by anharmonic lattice fluctuations which is consistent with the enhanced electronic screening and stronger lattice fluctuations expected from the more delocalized metavalent bonding.

The much larger LO-TO splitting in rs-BeO does not originate primarily from the Born effective charges, which differ only moderately between the two phases ($Z^* \sim 2.21$ for rs-BeO vs. 1.78 for zb-BeO). Rather, the splitting scales with $(Z^*)^2$ and the dielectric response. Using our calculated parameters, 
$\omega_{\mathrm{LO}}^2-\omega_{\mathrm{TO}}^2 \approx 1.79~\mathrm{THz}^2$ for rs-BeO and $\approx 0.48~\mathrm{THz}^2$ for zb-BeO. The directly computed frequency difference $\Delta \omega$ is $26.60~\mathrm{THz}$ for rs-BeO and $10.77~\mathrm{THz}$ for zb-BeO, consistent with their respective dielectric properties. 
However, the dramatic frequency gap in rs-BeO is amplified by its extremely soft TO mode ($\omega_{\mathrm{TO}} = 3.46~\mathrm{THz}$ vs. $22.25~\mathrm{THz}$ in zb-BeO), while the LO frequencies remain comparable ($\sim 30~\mathrm{THz}$ vs. $\sim 33~\mathrm{THz}$). This TO softening is a direct manifestation of the metavalent-like bonding character.

\begin{figure*}
\includegraphics[width=2.0\columnwidth]{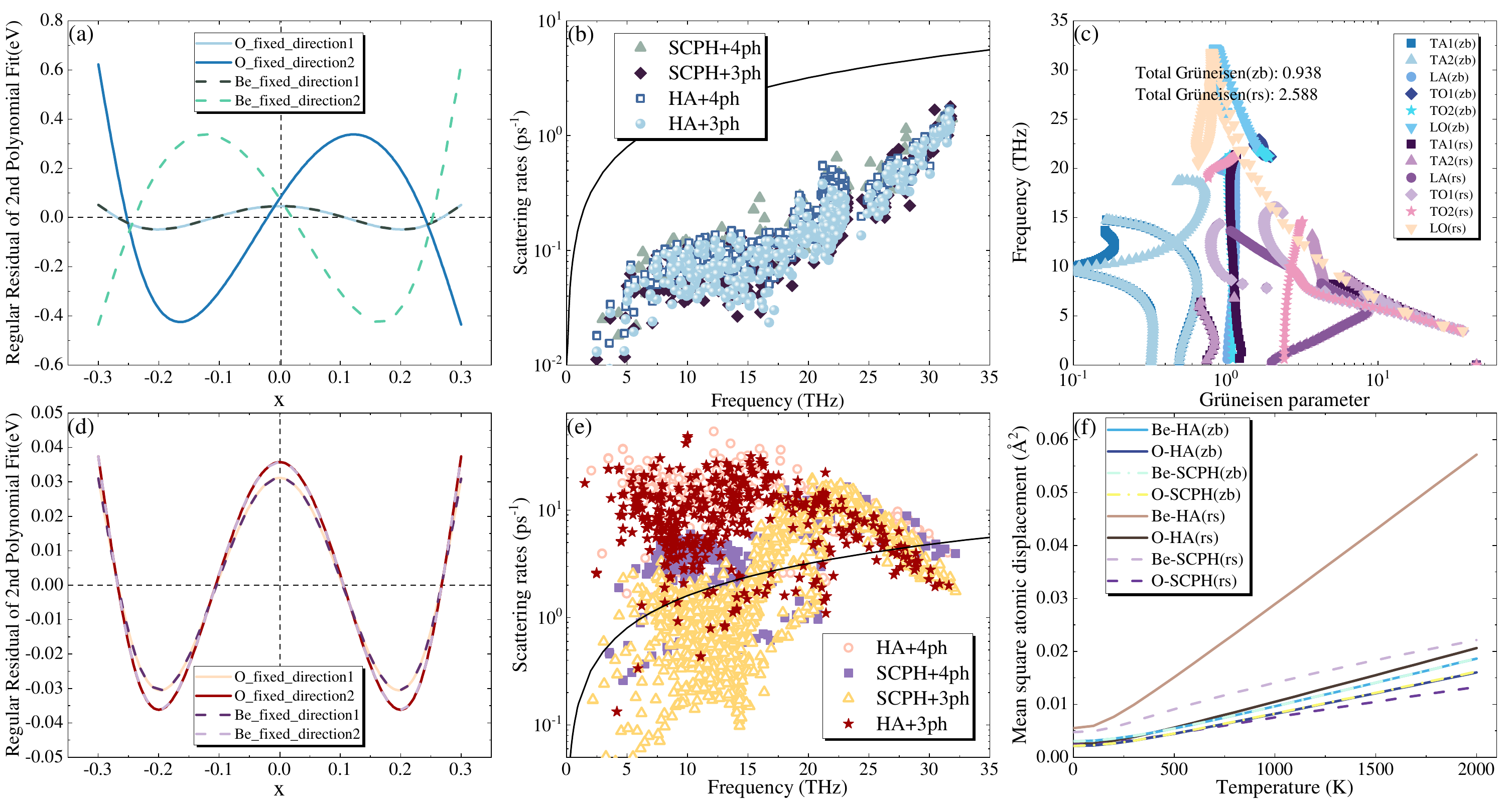}
\caption{Comprehensive analysis of anharmonic lattice dynamics in zb-BeO and rs-BeO. (a, d) Regular residuals from second-order polynomial fitting to the potential energy curves along two directions perpendicular to the Be-O bond for zb-BeO and rs-BeO, respectively. (b, e) Phonon-phonon scattering rates for zb-BeO and rs-BeO. The solid black line indicates the Ioffe-Regel limit (1/$\tau$=$\omega$/2$\pi$)~\cite{PhysRevX.12.041011}. (c) Mode-resolved Gr\"{u}neisen parameters at 300 K for transverse acoustic (TA), longitudinal acoustic (LA), transverse optical (TO), and longitudinal optical (LO) branches. The overall Gr\"{u}neisen parameters are 0.94 for zb-BeO and 2.59 for rs-BeO. (f) Mean square atomic displacements (MSDs) of Be and O atoms in the HA and SCPH treatments.
\label{fig6}}
\end{figure*}

Fig.~\ref{fig6} provides a comprehensive analysis of the anharmonic lattice behavior, building on the bonding and phonon results presented in Figs.~\ref{fig1}-\ref{fig5}. Figs.~\ref{fig6}(a) and (d) show the regular residual analysis (RRA), which quantifies higher-order anharmonic contributions along two directions perpendicular to the bond axis. In zb-BeO, the residual patterns are direction-dependent: along direction 1 (0, 0.7071, -0.7071) the curve exhibits a ``W''-like shape indicative of fourth-order contributions, while along direction 2 (-0.8165, 0.4082, 0.4082) the ``$\backsim$''-like form reflects dominant third-order anharmonicity.

This directional contrast indicates anisotropic anharmonic behavior with limited overall quartic influence. In rs-BeO, by contrast, consistent ``W''-shaped residuals appear along both directions, pointing to strong and isotropic quartic anharmonicity. The dominance of fourth-order terms in shaping the potential energy surface is consistent with the long-range interactions and flattened energy landscape identified from IFC analysis.


The phonon scattering rates are shown in Fig.~\ref{fig6}(b) for zb-BeO and Fig.~\ref{fig6}(e) for rs-BeO. Without renormalization, rs-BeO exhibits much larger scattering rates than zb-BeO across most of the spectrum, with the difference becoming most pronounced at low frequencies. Several phonon modes in rs-BeO approach the Ioffe–Regel limit (1/$\tau$=$\omega$/2$\pi$), implying a loss of well-defined quasiparticle character. Upon including SCPH renormalization, the scattering rates in rs-BeO decrease substantially by nearly an order of magnitude in the intermediate-frequency range, indicating that renormalization suppresses excessive scattering and restores phonon coherence. In contrast, the scattering rates in zb-BeO remain essentially unchanged after SCPH correction, confirming that anharmonic effects are intrinsically weak in this phase.


Fig.~\ref{fig6}(c) presents the branch-resolved Grüneisen parameters at 300 K. The overall value reaches 2.59 for rs-BeO, compared with only 0.94 for zb-BeO. The large value in rs-BeO exceeds those of many known metavalent materials and is distributed broadly across different phonon branches, reflecting strong sensitivity of phonon frequencies to lattice strain and indicating pervasive anharmonic interactions.


The temperature dependence of the atomic mean square displacements (MSDs) is shown in Fig.~\ref{fig6}(f). In zb-BeO, the harmonic approximation (HA) and SCPH results nearly overlap for both Be and O atoms, indicating that thermal vibrations are well described within the harmonic picture. In rs-BeO, the situation is markedly different: the HA results deviate significantly from the SCPH predictions at elevated temperatures, particularly for Be atoms, whose displacement amplitudes are approximately three times larger than those in zb-BeO. The SCPH method yields reduced MSDs compared to the HA, indicating that quartic anharmonicity hardens the phonon modes and limits the growth of atomic vibrations. This behavior is further supported by the potential energy surfaces of representative phonon modes (Fig. S13 in the Supplemental Material), which reveal a pronounced quartic contribution in rs-BeO compared to the nearly harmonic character in zb-BeO. Taken together, these results demonstrate that rs-BeO is governed by strong quartic anharmonicity originating from its metavalent-like bonding nature, whereas zb-BeO maintains weak anharmonicity characteristic of a conventional covalent system.

Having established a reliable lattice dynamics model, we now investigate $\kappa_L$ of zb-BeO and rs-BeO using progressively higher levels of theory, as shown in Figs.~\ref{fig5}(b) and (d). Our approach systematically compares different combinations of 3ph and 4ph scattering with and without anharmonic phonon renormalization (APRN). In zb-BeO, phonon renormalization has only a modest influence on thermal transport. At room temperature, including APRN reduces the 3ph-limited $\kappa_L$ from $\sim$520 to $\sim$485 $\mathrm{W~m^{-1}~K^{-1}}$, a decrease of less than 7\%. Further incorporation of 4ph scattering ($\kappa^{\rm 3,4ph}_{\rm SCPH}$) yields a 27\% reduction, giving a value of $\sim$353 $\mathrm{W~m^{-1}~K^{-1}}$ that surpasses the vast majority of known semiconductors. This high $\kappa_L$ reflects the stiff covalent bonding in zb-BeO, which produces weak anharmonicity that is consistent with the nearly temperature-invariant phonon dispersions and the small $\gamma$.


The situation in rs-BeO is strikingly different. Phonon renormalization leads to a substantial increase in $\kappa_L$. Within the 3ph picture, APRN raises the room-temperature $\kappa_L$ from 12 to 61 $\mathrm{W~m^{-1}~K^{-1}}$ (408.3\% enhancement). When 4ph scattering is further included, APRN increases $\kappa_L$ from 5 to 23 $\mathrm{W~m^{-1}~K^{-1}}$ (360\% enhancement). The additional coherent contribution slightly raises the total $\kappa_L$ to 24~W~m$^{-1}$~K$^{-1}$. This counterintuitive behavior arises because renormalization hardens the soft phonon modes, reduces the scattering phase space, and extends phonon lifetimes. Concurrently, the APRN effects significantly weaken the temperature dependence of $\kappa_L$. 

Notably, $\kappa^{\rm 3ph}_{\rm SCPH}$ even exhibits an upward trend at elevated temperatures. Four-phonon scattering exerts a pronounced suppressive effect on $\kappa_L$ in rs-BeO, reducing it by 58.3\% and 62.3\% at room temperature within the HA and SCPH frameworks, respectively. The inclusion of coherent transport has a negligible effect on $\kappa_L$ for zb-BeO over the entire temperature range and contributes only modestly to rs-BeO at high temperatures. Taking all contributions into account, the predicted room-temperature $\kappa_L$ of rs-BeO is 24 $\mathrm{W~m^{-1}~K^{-1}}$ which is more than an order of magnitude lower than that of zb-BeO.
These results demonstrate that both APRN and four-phonon scattering are essential for rs-BeO: neglecting APRN underestimates $\kappa_L$, whereas neglecting four-phonon scattering overestimates it. Their partial cancellation at room temperature is accidental and does not justify omitting either effect~\cite{xia2018revisiting,wang2024revisiting}.

Our study reveals a pronounced anomaly: although $\kappa_L$ generally increases with density, rs-BeO ($\rho = 3.60~\mathrm{g\,cm^{-3}}$) lies more than tenfold below the less dense zb-BeO ($\rho = 3.12~\mathrm{g\,cm^{-3}}$). The breakdown of the density–$\kappa_L$ correlation cannot be explained by structural parameters alone and instead originates in the contrasting bonding characters of the two phases (Figs.~\ref{fig2} and \ref{fig3}), with similar behavior expected in other materials near the covalent–ionic boundary.

In summary, our controlled comparison of two chemically identical BeO polymorphs establishes that bonding character, rather than density, governs lattice thermal transport: despite being denser, rs-BeO exhibits a $\kappa_L$ more than an order of magnitude below that of zb-BeO, because its metavalent-like bonding softens the lattice and generates giant quartic anharmonicity. Quantitative modeling of such systems requires the explicit inclusion of four-phonon scattering, temperature-dependent phonon renormalization, and coherent transport. At 300 K, neglecting renormalization underestimates $\kappa_L$ by a factor of 4.6. Notably, these metavalent-like bonding signatures survive in an ultra-wide-bandgap insulator, showing that electronic conductivity is not a prerequisite for this bonding character. On this basis we propose three practical indicators for identifying metavalent-like bonding: an NaCl-type structure, a Gr\"{u}neisen parameter above 2, and mixed bonding-antibonding states near Fermi level accompanied by spatially delocalized orbitals extending beyond nearest neighbors. Together these provide a diagnostic route toward strongly anharmonic thermoelectric and phase-change materials.

\section{Additional information}
The Supplementary Material contains: crystal structures and lattice parameters; phonon dispersions under varying pressures (0-300 GPa) and temperatures (0-2000 K); temperature-dependent thermodynamic properties (Helmholtz free energy, internal energy, entropy, and heat capacity); phonon dispersions, charge density distortions, and normalized IFC traces calculated with PBE and PBEsol functionals; electronic band structures and projected density of states; phonon group velocities; potential energy surfaces; and decomposed three- and four-phonon scattering rates for both zb-BeO and rs-BeO. It also includes AIMD total-energy trajectories and partial radial distribution functions at 0 and 200~GPa, which are used to assess the finite-temperature dynamical stability of rs-BeO. The data supporting Fig.~\ref{fig1} are tabulated with corresponding references.

\section{Acknowledgements}
The authors appreciate discussions with Prof. Ruyue Cao at the Chinese Academy of Sciences about the experimental synthesis of rs-BeO.
%
We acknowledge the support from the National Natural Science Foundation of China 
(No.52595632,  
No.12374038, 
and No.12674045)  
and Xi'an Jiaotong University-China Mobile Communications Group Co.,Ltd. Joint Institute. 
This work was sponsored by the Key Research and Development Program of China (No.2023YFB4604100). 
%
X. Yang acknowledges support from the Fundamental Research Fund for the Central Universities (Grant No. 2025CDJ-IAISYB-035) and the New Chongqing Youth Innovative Talent Project (Grant No. CSTB2025YITP-QCRCX006).
We also acknowledge the support by HPC Platform, Xi’an Jiaotong University. 


\section{Author contributions}
Z.G. conceived the research idea. X.L. performed the calculations, created the visualizations, and wrote the original manuscript draft. Z.G., X.Y., T.L., X.D. and S.Y. supervised the research. All authors participated in discussions, provided critical feedback, and reviewed the manuscript.

\section{Competing interests}
The authors declare no competing financial interest.

\renewcommand{\bibsection}{\section*{References}}
\bibliography{References}

@article{PhysRevX.12.041011,
  title = {Wigner Formulation of Thermal Transport in Solids},
  author = {Simoncelli, Michele and Marzari, Nicola and Mauri, Francesco},
  journal = {Phys. Rev. X},
  volume = {12},
  issue = {4},
  pages = {041011},
  numpages = {39},
  year = {2022},
  month = {Oct},
  publisher = {American Physical Society},
  doi = {10.1103/PhysRevX.12.041011},
  url = {https://link.aps.org/doi/10.1103/PhysRevX.12.041011}
}

@article{hao2026llm,
  title={{LLM}-driven discovery for carbon allotropes with bond-network entropy},
  author={Hao, Yuzhou and Liu, Yujie and Li, Xuejie and Lookman, Turab and Ding, Xiangdong and Sun, Jun and Gao, Zhibin},
  journal={Appl. Phys. Lett.},
  volume={128},
  number={10},
  year={2026},
  url = {https://doi.org/10.1063/5.0314279},
  publisher={AIP Publishing}
}

@article{MALAKKAL201779,
title = {Thermal conductivity of wurtzite and zinc blende cubic phases of {BeO} from ab initio calculations},
journal = {Solid State Sci.},
volume = {65},
pages = {79-87},
year = {2017},
issn = {1293-2558},
doi = {https://doi.org/10.1016/j.solidstatesciences.2017.01.005},
url = {https://www.sciencedirect.com/science/article/pii/S1293255816307610},
author = {Linu Malakkal and Barbara Szpunar and Ravi Kiran Siripurapu and Juan Carlos Zuniga and Jerzy A. Szpunar},
}

@article{10.1063/1.3075814,
    author = {Duman, S. and Sütlü, A. and Bağcı, S. and Tütüncü, H. M. and Srivastava, G. P.},
    title = {Structural, elastic, electronic, and phonon properties of zinc-blende and wurtzite $\textrm{BeO}$},
    journal = {J. Appl. Phys.},
    volume = {105},
    number = {3},
    pages = {033719},
    year = {2009},
    month = {02},
    issn = {0021-8979},
    doi = {10.1063/1.3075814},
    url = {https://doi.org/10.1063/1.3075814},
}

@article{Ivanovskii2009,
  author = {Ivanovskii, A. L. and Shein, I. R. and Makurin, Yu. N. and
    Kiiko, V. S. and Gorbunova, M. A.},
  title = {Electronic Structure and Properties of Beryllium Oxide},
  journal = {Inorg. Mater.},
  volume = {45},
  number = {3},
  pages = {223-234},
  year = {2009},
  url = {https://doi.org/10.1134/S0020168509030017}
}

@article{Sorokin2006,
  author = {Sorokin, P. B. and Fedorov, A. S. and Chernozatonskiĭ, L.
    A.},
  title = {Structure and Properties of {BeO} Nanotubes},
  journal = {Phys. Solid State},
  volume = {48},
  number = {2},
  pages = {398-401},
  year = {2006},
  url = {https://doi.org/10.1134/S106378340602034X}
}

@article{PhysRevB.76.085407,
  title = {Structural, elastic, and electronic properties of {SiC}, {BN}, and {BeO} nanotubes},
  author = {Baumeier, Bj\"orn and Kr\"uger, Peter and Pollmann, Johannes},
  journal = {Phys. Rev. B},
  volume = {76},
  issue = {8},
  pages = {085407},
  numpages = {10},
  year = {2007},
  month = {Aug},
  publisher = {American Physical Society},
  doi = {10.1103/PhysRevB.76.085407},
  url = {https://link.aps.org/doi/10.1103/PhysRevB.76.085407}
}

@article{10.1063/1.336756,
    author = {Hazen, R. M. and Finger, L. W.},
    title = {High‐pressure and high‐temperature crystal chemistry of beryllium oxide},
    journal = {J. Appl. Phys.},
    volume = {59},
    number = {11},
    pages = {3728-3733},
    year = {1986},
    month = {06},
    issn = {0021-8979},
    doi = {10.1063/1.336756},
    url = {https://doi.org/10.1063/1.336756},
}

@article{10.1063/1.1659844,
    author = {Slack, Glen A. and Austerman, S. B.},
    title = {Thermal Conductivity of {BeO} Single Crystals},
    journal = {J. Appl. Phys.},
    volume = {42},
    number = {12},
    pages = {4713-4717},
    year = {1971},
    month = {11},
    issn = {0021-8979},
    doi = {10.1063/1.1659844},
    url = {https://doi.org/10.1063/1.1659844},
}

@article{Vidal-Valat:a26524,
author = "Vidal-Valat, G. and Vidal, J. P. and Kurki-Suonio, K. and Kurki-Suonio, R.",
title = "{Multipole analysis of X-ray diffraction data on BeO}",
journal = "Acta Crystallogr. A",
year = "1987",
volume = "43",
number = "4",
pages = "540--550",
month = "Jul",
doi = {10.1107/S0108767387099057},
url = {https://doi.org/10.1107/S0108767387099057},
}

@article{Qi-Li_2008,
doi = {10.1088/1674-1056/17/4/031},
url = {https://dx.doi.org/10.1088/1674-1056/17/4/031},
year = {2008},
month = {apr},
publisher = {},
volume = {17},
number = {4},
pages = {1341},
author = {Zhang Qi-Li and Zhang Ping and Song Hai-Feng and Liu Hai-Feng},
title = {Mean-field potential calculations of high-pressure equation of state for {BeO}},
journal = {Chin. Phys. B},
}

@article{ROESSLER1969157,
title = {Electronic spectrum of crystalline beryllium oxide},
journal = {J. Phys. Chem. Solids},
volume = {30},
number = {1},
pages = {157-167},
year = {1969},
issn = {0022-3697},
doi = {https://doi.org/10.1016/0022-3697(69)90348-5},
url = {https://www.sciencedirect.com/science/article/pii/0022369769903485},
author = {D.M. Roessler and W.C. Walker and Eugene Loh},
}

@article{Wdowik_2010,
doi = {10.1088/0953-8984/22/4/045404},
url = {https://dx.doi.org/10.1088/0953-8984/22/4/045404},
year = {2010},
month = {jan},
publisher = {},
volume = {22},
number = {4},
pages = {045404},
author = {Wdowik, Urszula D},
title = {Structural stability and thermal properties of {BeO} from the quasiharmonic approximation},
journal = {J. Phys.: Condens. Matter},
}

@article{10.1063/1.3359706,
    author = {Sahariah, Munima B. and Ghosh, Subhradip},
    title = {Dynamical stability and phase transition of {BeO} under pressure},
    journal = {J. Appl. Phys.},
    volume = {107},
    number = {8},
    pages = {083520},
    year = {2010},
    month = {04},
    issn = {0021-8979},
    doi = {10.1063/1.3359706},
    url = {https://doi.org/10.1063/1.3359706},
}

@article{https://doi.org/10.1111/j.1151-2916.1966.tb15389.x,
author = {BENTLE, G. G.},
title = {Elastic Constants of Single-Crystal {BeO} at Room Temperature},
journal = {J. Am. Ceram. Soc.},
volume = {49},
number = {3},
pages = {125-128},
doi = {https://doi.org/10.1111/j.1151-2916.1966.tb15389.x},
url = {https://ceramics.onlinelibrary.wiley.com/doi/abs/10.1111/j.1151-2916.1966.tb15389.x},
year = {1966}
}

@article{10.1063/1.1709787,
    author = {Cline, Carl F. and Dunegan, Harold L. and Henderson, Glenn W.},
    title = {Elastic Constants of Hexagonal {BeO, ZnS, and CdSe}},
    journal = {J. Appl. Phys.},
    volume = {38},
    number = {4},
    pages = {1944-1948},
    year = {1967},
    month = {03},
    issn = {0021-8979},
    doi = {10.1063/1.1709787},
    url = {https://doi.org/10.1063/1.1709787},
}

@article{doi:10.1021/acsomega.9b00174,
author = {Wei, Jie and Zhou, Wei and Li, Song and Shen, Pei and Ren, Shuai and Hu, Alice and Zhou, Wenzhong},
title = {Modified Embedded Atom Method Potential for Modeling the Thermodynamic Properties of High Thermal Conductivity Beryllium Oxide},
journal = {ACS Omega},
volume = {4},
number = {4},
pages = {6339-6346},
year = {2019},
doi = {10.1021/acsomega.9b00174},
URL = {    
        https://doi.org/10.1021/acsomega.9b00174
},
}

@article{PhysRevB.77.224303,
  title = {Lattice dynamics of beryllium oxide: Inelastic {X}-ray scattering and ab initio calculations},
  author = {Bosak, Alexey and Schmalzl, Karin and Krisch, Michael and van Beek, Wouter and Kolobanov, Vitaly},
  journal = {Phys. Rev. B},
  volume = {77},
  issue = {22},
  pages = {224303},
  numpages = {7},
  year = {2008},
  month = {Jun},
  publisher = {American Physical Society},
  doi = {10.1103/PhysRevB.77.224303},
  url = {https://link.aps.org/doi/10.1103/PhysRevB.77.224303}
}

@article{Milman_2001,
doi = {10.1088/0953-8984/13/2/302},
url = {https://dx.doi.org/10.1088/0953-8984/13/2/302},
year = {2001},
month = {jan},
publisher = {},
volume = {13},
number = {2},
pages = {241},
author = {V Milman and M C Warren},
title = {Elasticity of hexagonal {BeO}},
journal = {J. Phys.: Condens. Matter},
}

@article{xia2018revisiting,
  title={{Revisiting lattice thermal transport in PbTe: The crucial role of quartic anharmonicity}},
  author={Xia, Yi},
  journal={Appl. Phys. Lett.},
  volume={113},
  number={7},
  year={2018},
  url={https://doi.org/10.1063/1.5040887},
  publisher={AIP Publishing}
}

@article{wang2024revisiting,
  title={{Revisiting lattice thermal conductivity of CsCl: The crucial role of quartic anharmonicity}},
  author={Wang, Xiaoying and Feng, Minxuan and Xia, Yi and Sun, Jun and Ding, Xiangdong and Li, Baowen and Gao, Zhibin},
  journal={Appl. Phys. Lett.},
  volume={124},
  number={17},
  year={2024},
  url={https://doi.org/10.1063/5.0201393},
  publisher={AIP Publishing}
}

@article{Cao2024,
  author = {Cao, Ruyue and Yang, Qiao-Lin and Deng, Hui-Xiong and Wei,
    Su-Huai and Robertson, John and Luo, Jun-Wei},
  title = {Softening of the Optical Phonon by Reduced Interatomic
    Bonding Strength Without Depolarization},
  journal = {Nature},
  volume = {634},
  number = {8036},
  pages = {1080-1085},
  year = {2024},
  url = {https://doi.org/10.1038/s41586-024-08099-0}
}

@article{li2026four,
  title={{Four-phonon scattering and coherent heat transport in ultrawide-bandgap SrSnO$_3$}},
  author={Li, Xuejie and Wang, Xiaoying and Yue, Shengying and Lookman, Turab and Ding, Xiangdong and Sun, Jun and Gao, Zhibin},
  journal={Appl. Phys. Lett.},
  volume={129},
  number={8},
  year={2026},
  url={https://doi.org/10.1063/5.0348652},
  publisher={AIP Publishing}
}

@article{wang2023role,
  title = {{Role of high-order anharmonicity and off-diagonal terms in thermal conductivity: A case study of multiphase ${\mathrm{CsPbBr}}_{3}$}},
  author = {Wang, Xiaoying and Gao, Zhibin and Zhu, Guimei and Ren, Jie and Hu, Lei and Sun, Jun and Ding, Xiangdong and Xia, Yi and Li, Baowen},
  journal = {Phys. Rev. B},
  volume = {107},
  issue = {21},
  pages = {214308},
  numpages = {8},
  year = {2023},
  month = {Jun},
  publisher = {American Physical Society},
  doi = {10.1103/PhysRevB.107.214308},
  url = {https://link.aps.org/doi/10.1103/PhysRevB.107.214308}
}

@article{hao2025copper,
  title = {{Copper delocalization leads to ultralow thermal conductivity in chalcohalide ${\mathrm{CuBiSeCl}}_{2}$}},
  author = {Hao, Yuzhou and Che, Junwei and Wang, Xiaoying and Li, Xuejie and Lookman, Turab and Sun, Jun and Ding, Xiangdong and Gao, Zhibin},
  journal = {Phys. Rev. B},
  volume = {111},
  issue = {19},
  pages = {195207},
  numpages = {10},
  year = {2025},
  month = {May},
  publisher = {American Physical Society},
  doi = {10.1103/PhysRevB.111.195207},
  url = {https://link.aps.org/doi/10.1103/PhysRevB.111.195207}
}

@article{https://doi.org/10.1002/adma.201803777,
author = {Wuttig, Matthias and Deringer, Volker L. and Gonze, Xavier and Bichara, Christophe and Raty, Jean-Yves},
title = {Incipient Metals: Functional Materials with a Unique Bonding Mechanism},
journal = {Adv. Mater.},
volume = {30},
number = {51},
pages = {1803777},
doi = {https://doi.org/10.1002/adma.201803777},
url = {https://advanced.onlinelibrary.wiley.com/doi/abs/10.1002/adma.201803777},
year = {2018}
}

@article{Shportko2008,
  author = {Shportko, Kostiantyn and Kremers, Stephan and Woda, Michael
    and Lencer, Dominic and Robertson, John and Wuttig, Matthias},
  title = {Resonant Bonding in Crystalline Phase-Change Materials},
  journal = {Nat. Mater.},
  volume = {7},
  number = {8},
  pages = {653-658},
  year = {2008},
  url = {https://doi.org/10.1038/nmat2226}
}

@article{https://doi.org/10.1002/pssb.200982010,
author = {Wuttig, Matthias},
title = {Phase change materials: The importance of resonance bonding},
journal = {Phys. Status Solidi B},
volume = {246},
number = {8},
pages = {1820-1825},
doi = {https://doi.org/10.1002/pssb.200982010},
url = {https://onlinelibrary.wiley.com/doi/abs/10.1002/pssb.200982010},
year = {2009}
}

@article{doi:10.1021/acs.nanolett.9b03435,
author = {Raghuwanshi, Mohit and Cojocaru-Mir{\'e}din, Oana and Wuttig, Matthias},
title = {Investigating Bond Rupture in Resonantly Bonded Solids by Field Evaporation of Carbon Nanotubes},
journal = {Nano Lett.},
volume = {20},
number = {1},
pages = {116-121},
year = {2020},
doi = {10.1021/acs.nanolett.9b03435},
URL = {     
        https://doi.org/10.1021/acs.nanolett.9b03435
},
}

@article{https://doi.org/10.1002/adma.201806280,
author = {Raty, Jean-Yves and Schumacher, Mathias and Golub, Pavlo and Deringer, Volker L. and Gatti, Carlo and Wuttig, Matthias},
title = {A Quantum-Mechanical Map for Bonding and Properties in Solids},
journal = {Adv. Mater.},
volume = {31},
number = {3},
pages = {1806280},
doi = {https://doi.org/10.1002/adma.201806280},
url = {https://advanced.onlinelibrary.wiley.com/doi/abs/10.1002/adma.201806280},
year = {2019}
}

@article{https://doi.org/10.1002/adma.202208485,
author = {Wuttig, Matthias and Schön, Carl-Friedrich and Lötfering, Jakob and Golub, Pavlo and Gatti, Carlo and Raty, Jean-Yves},
title = {Revisiting the Nature of Chemical Bonding in Chalcogenides to Explain and Design their Properties},
journal = {Adv. Mater.},
volume = {35},
number = {20},
pages = {2208485},
doi = {https://doi.org/10.1002/adma.202208485},
url = {https://advanced.onlinelibrary.wiley.com/doi/abs/10.1002/adma.202208485},
year = {2023}
}

@article{https://doi.org/10.1002/advs.202308578,
author = {Wuttig, Matthias and Schön, Carl-Friedrich and Kim, Dasol and Golub, Pavlo and Gatti, Carlo and Raty, Jean-Yves and Kooi, Bart J. and Pendás, Ángel Martín and Arora, Raagya and Waghmare, Umesh},
title = {Metavalent or Hypervalent Bonding: Is There a Chance for Reconciliation?},
journal = {Adv. Sci.},
volume = {11},
number = {6},
pages = {2308578},
doi = {https://doi.org/10.1002/advs.202308578},
url = {https://advanced.onlinelibrary.wiley.com/doi/abs/10.1002/advs.202308578},
year = {2024}
}

@article{https://doi.org/10.1002/adma.201801787,
author = {Cagnoni, Matteo and Führen, Daniel and Wuttig, Matthias},
title = {Thermoelectric Performance of {IV–VI} Compounds with Octahedral-Like Coordination: A Chemical-Bonding Perspective},
journal = {Adv. Mater.},
volume = {30},
number = {33},
pages = {1801787},
doi = {https://doi.org/10.1002/adma.201801787},
url = {https://advanced.onlinelibrary.wiley.com/doi/abs/10.1002/adma.201801787},
year = {2018}
}

@article{https://doi.org/10.1002/adfm.201904862,
author = {Yu, Yuan and Cagnoni, Matteo and Cojocaru-Mirédin, Oana and Wuttig, Matthias},
title = {Chalcogenide Thermoelectrics Empowered by an Unconventional Bonding Mechanism},
journal = {Adv. Funct. Mater.},
volume = {30},
number = {8},
pages = {1904862},
doi = {https://doi.org/10.1002/adfm.201904862},
url = {https://advanced.onlinelibrary.wiley.com/doi/abs/10.1002/adfm.201904862},
year = {2020}
}

@article{PhysRevB.8.660,
  title = {Effects of Resonance Bonding on the Properties of Crystalline and Amorphous Semiconductors},
  author = {Lucovsky, G. and White, R. M.},
  journal = {Phys. Rev. B},
  volume = {8},
  issue = {2},
  pages = {660--667},
  numpages = {0},
  year = {1973},
  month = {Jul},
  publisher = {American Physical Society},
  doi = {10.1103/PhysRevB.8.660},
  url = {https://link.aps.org/doi/10.1103/PhysRevB.8.660}
}

@article{PBLittlewood_1980,
doi = {10.1088/0022-3719/13/26/009},
url = {https://dx.doi.org/10.1088/0022-3719/13/26/009},
year = {1980},
month = {sep},
publisher = {},
volume = {13},
number = {26},
pages = {4855},
author = {P B Littlewood},
title = {The crystal structure of {IV-VI} compounds. {I}. Classification and description},
journal = {J. Phys. C: Solid State Phys.},
}

@article{doi:10.1126/science.aad8688,
author = {Kamran Behnia },
title = {Finding merit in dividing neighbors},
journal = {Science},
volume = {351},
number = {6269},
pages = {124-124},
year = {2016},
doi = {10.1126/science.aad8688},
URL = {https://www.science.org/doi/abs/10.1126/science.aad8688},
}

@article{10.1147/rd.83.0215,
author = {Cohen, Morrel H. and Falicov, L. M. and Golin, Stuart},
title = {Crystal chemistry and band structures of the group {V} semimetals and the {IV-VI} semiconductors},
year = {1964},
issue_date = {July 1964},
publisher = {IBM Corp.},
address = {USA},
volume = {8},
number = {3},
issn = {0018-8646},
url = {https://doi.org/10.1147/rd.83.0215},
doi = {10.1147/rd.83.0215},
journal = {IBM J. Res. Dev.},
month = jul,
pages = {215–227},
numpages = {13}
}

@article{CHATTOPADHYAY1986879,
title = {Neutron diffraction study of the structural phase transition in {SnS} and {SnSe}},
journal = {J. Phys. Chem. Solids},
volume = {47},
number = {9},
pages = {879-885},
year = {1986},
issn = {0022-3697},
doi = {https://doi.org/10.1016/0022-3697(86)90059-4},
url = {https://www.sciencedirect.com/science/article/pii/0022369786900594},
author = {T. Chattopadhyay and J. Pannetier and H.G. {Von Schnering}},
}

@article{Lencer2008,
  author = {Lencer, Dominic and Salinga, Martin and Grabowski, Blazej
    and Hickel, Tilmann and Neugebauer, Jörg and Wuttig, Matthias},
  title = {A Map for Phase-Change Materials},
  journal = {Nat. Mater.},
  volume = {7},
  number = {12},
  pages = {972-977},
  year = {2008},
  url = {https://doi.org/10.1038/nmat2330}
}

@article{Lee2014,
  author = {Lee, Sangyeop and Esfarjani, Keivan and Luo, Tengfei and
    Zhou, Jiawei and Tian, Zhiting and Chen, Gang},
  title = {Resonant Bonding Leads to Low Lattice Thermal Conductivity},
  journal = {Nat. Commun.},
  volume = {5},
  number = {1},
  pages = {3525},
  year = {2014},
  url = {https://doi.org/10.1038/ncomms4525}
}

@article{PhysRevLett.107.175503,
  title = {Anomalous Lattice Dynamics near the Ferroelectric Instability in {PbTe}},
  author = {Zhang, Yi and Ke, Xuezhi and Kent, Paul R. C. and Yang, Jihui and Chen, Changfeng},
  journal = {Phys. Rev. Lett.},
  volume = {107},
  issue = {17},
  pages = {175503},
  numpages = {4},
  year = {2011},
  month = {Oct},
  publisher = {American Physical Society},
  doi = {10.1103/PhysRevLett.107.175503},
  url = {https://link.aps.org/doi/10.1103/PhysRevLett.107.175503}
}

@article{YUE201889,
title = {Ultralow thermal conductivity in a two-dimensional material due to surface-enhanced resonant bonding},
journal = {Mater. Today Phys.},
volume = {7},
pages = {89-95},
year = {2018},
issn = {2542-5293},
doi = {https://doi.org/10.1016/j.mtphys.2018.11.005},
url = {https://www.sciencedirect.com/science/article/pii/S2542529318301317},
author = {Sheng-Ying Yue and Tashi Xu and Bolin Liao},
}

@article{
doi:10.1126/sciadv.abg1449,
author = {Jingxuan Ding  and Tyson Lanigan-Atkins  and Mario Calderón-Cueva  and Arnab Banerjee  and Douglas L. Abernathy  and Ayman Said  and Alexandra Zevalkink  and Olivier Delaire },
title = {Soft anharmonic phonons and ultralow thermal conductivity in $\textrm{Mg}_3\textrm{(Sb,Bi)}_2$ thermoelectrics},
journal = {Sci. Adv.},
volume = {7},
number = {21},
pages = {eabg1449},
year = {2021},
doi = {10.1126/sciadv.abg1449},
URL = {https://www.science.org/doi/abs/10.1126/sciadv.abg1449},
}

@article{10.1063/5.0271507,
    author = {Wang, Yue and Liu, Anping and Yang, Xiaolong and Zhou, Xiaoyuan},
    title = {Anomalous thermal conductivity anisotropy in bulk hexagonal AlN induced by structural phase transition},
    journal = {Appl. Phys. Lett.},
    volume = {126},
    number = {20},
    pages = {202207},
    year = {2025},
    month = {05},
    issn = {0003-6951},
    doi = {10.1063/5.0271507},
    url = {https://doi.org/10.1063/5.0271507},
}

@article{https://doi.org/10.1002/jcc.26353,
author = {Nelson, Ryky and Ertural, Christina and George, Janine and Deringer, Volker L. and Hautier, Geoffroy and Dronskowski, Richard},
title = {LOBSTER: Local orbital projections, atomic charges, and chemical-bonding analysis from projector-augmented-wave-based density-functional theory},
journal = {Journal of Computational Chemistry},
volume = {41},
number = {21},
pages = {1931-1940},
doi = {https://doi.org/10.1002/jcc.26353},
url = {https://onlinelibrary.wiley.com/doi/abs/10.1002/jcc.26353},
year = {2020}
}

@article{Yim2015,
  author = {Yim, Kanghoon and Yong, Youn and Lee, Joohee and Lee,
    Kyuhyun and Nahm, Ho-Hyun and Yoo, Jiho and Lee, Chanhee and Seong
    Hwang, Cheol and Han, Seungwu},
  title = {Novel High-$\kappa$ Dielectrics for Next-Generation Electronic
    Devices Screened by Automated Ab Initio Calculations},
  journal = {NPG Asia Mater.},
  volume = {7},
  number = {6},
  pages = {e190-e190},
  year = {2015},
  url = {https://doi.org/10.1038/am.2015.57}
}

@article{https://doi.org/10.1002/pssb.200405247,
author = {Mori, Y. and Niiya, N. and Ukegawa, K. and Mizuno, T. and Takarabe, K. and Ruoff, A. L.},
title = {High-pressure {X}-ray structural study of {BeO} and {ZnO} to 200 {GPa}},
journal = {Phys. Stat. Sol. (b)},
volume = {241},
number = {14},
pages = {3198-3202},
doi = {https://doi.org/10.1002/pssb.200405247},
url = {https://onlinelibrary.wiley.com/doi/abs/10.1002/pssb.200405247},
year = {2004}
}

@article{PhysRevB.31.8332,
  title = {High-pressure {Raman} study of the optic-phonon modes in {BeO}},
  author = {Kourouklis, G. A. and Sood, A. K. and Hochheimer, H. D. and Jayaraman, A.},
  journal = {Phys. Rev. B},
  volume = {31},
  issue = {12},
  pages = {8332(R)--8334(R)},
  numpages = {0},
  year = {1985},
  month = {Jun},
  publisher = {American Physical Society},
  doi = {10.1103/PhysRevB.31.8332},
  url = {https://link.aps.org/doi/10.1103/PhysRevB.31.8332}
}

@article{10.1063/1.5040887,
    author = {Xia, Yi},
    title = {Revisiting lattice thermal transport in PbTe: The crucial role of quartic anharmonicity},
    journal = {Appl. Phys. Lett.},
    volume = {113},
    number = {7},
    pages = {073901},
    year = {2018},
    month = {08},
    issn = {0003-6951},
    doi = {10.1063/1.5040887},
    url = {https://doi.org/10.1063/1.5040887}
}

@article{10.1063/5.0201393,
    author = {Wang, Xiaoying and Feng, Minxuan and Xia, Yi and Sun, Jun and Ding, Xiangdong and Li, Baowen and Gao, Zhibin},
    title = {Revisiting lattice thermal conductivity of CsCl: The crucial role of quartic anharmonicity},
    journal = {Appl. Phys. Lett.},
    volume = {124},
    number = {17},
    pages = {172201},
    year = {2024},
    month = {04},
    issn = {0003-6951},
    doi = {10.1063/5.0201393},
    url = {https://doi.org/10.1063/5.0201393}
}







\end{document}